\documentclass[amsmath,amssymb,aps,prl,reprint,floatfix]{revtex4-1}

\usepackage{amssymb}
\usepackage{amsmath}
\usepackage[usenames,dvipsnames]{xcolor}
\usepackage[colorlinks=true,citecolor=blue,linkcolor=blue]{hyperref}
\usepackage{graphicx}
\usepackage{float}



\begin{document}
\title{Operating a multi-ion clock with dynamical decoupling}

\author{Nitzan Akerman}
\affiliation{Department of Physics of Complex Systems and AMOS, Weizmann Institute of Science,
Rehovot 7610001, Israel}

\author{Roee Ozeri}
\affiliation{Department of Physics of Complex Systems and AMOS, Weizmann Institute of Science,
Rehovot 7610001, Israel}

\begin{abstract}
We study and characterize a quasi-continuous dynamical decoupling (QCDD) scheme that effectively suppresses dominant frequency shifts in a multi-ion optical clock. Addressing the challenge of inhomogeneous frequency shifts in such systems, our scheme mitigates primary contributors, namely the electric quadrupole shift (QPS) and the linear Zeeman shift (LZS). Based on $^{88}$Sr$^+$ ions, we implement a QCDD scheme in linear chains of up to 7 ions and demonstrate a significant suppression of the shift by more than three orders of magnitude, leading to relative frequency inhomogeneity below $7\cdot10^{-17}$. Additionally, we evaluate the associated systematic shift arising from the radiofrequency (RF) drive used in the QCDD scheme, showing that, in the presented realization, its contribution to the systematic relative frequency uncertainty is below $10^{-17}$, with potential for further improvement. These results provide a promising avenue toward implementing multi-ion clocks exhibiting an order of magnitude or more improvement in stability while maintaining a similar high degree of accuracy to that of single-ion clocks.
\end{abstract}
\maketitle
Optical atomic clocks are a pinnacle achievement in quantum technology, delivering unparalleled precision with wide-ranging applications from fundamental science to technological advancements \cite{ludlow2015optical, safronova2018search}. Optical atomic transitions, as manifestations of accurate quantum oscillators, provide exceptional accuracy due to their high-quality factor and their natural indistinguishability, which will be practically impossible to achieve in macroscopic objects.

The trade-off to the high accuracy that atomic systems offer is their inherent quantum projection noise during measurement, which compromises their stability at the single-atom level. This entails a very long averaging time to reach the current clocks state-of-the-art accuracy at the $10^{-18}$ level \cite{brewer2019al+,huntemann2016single,huang2022liquid}. More importantly, long averaging times hinder applications that search for time-dependent signals\cite{derevianko2022fundamental,aharony2021constraining,dreissen2022improved}. Nevertheless, having $N$ identical atoms can be used to average projection noise faster and improve clock stability as $\sqrt{N}$ according to the standard quantum limit for independent particles. This methodology is pursued in neutral atoms in optical lattices and tweezers \cite{young2020half,aeppli2024clock,shaw2024multi}.

Challenges arise when considering multi-ion clock spectroscopy \cite{arnold2015prospects,keller2016evaluation,leibrandt2024prospects}, mostly due to inhomogeneous frequency shifts in Paul traps, which have largely impeded the realization of multi-ion clocks. One prominent example of an inhomogeneous shift in trapped ion systems is due to the electric field gradients of the trapping potential, which interact with the electric quadrupole moment of the clock transition states, introducing a tensorial QPS \cite{itano2000external,dube2005electric}. This shift becomes a crucial factor that compromises the accurate determination of the clock transition frequency. While traditional mitigation methods are successful in the single-ion case \cite{oskay2005measurement,margolis2004hertz, Schneider2005subHertz,lange2020coherent,chwalla2009absolute}, they are typically insufficient for multi-ion setups.

\begin{figure}[t]
    \centering
    \includegraphics[trim={0.5cm 2.9cm 1cm 0.8cm},clip,width=1\linewidth]{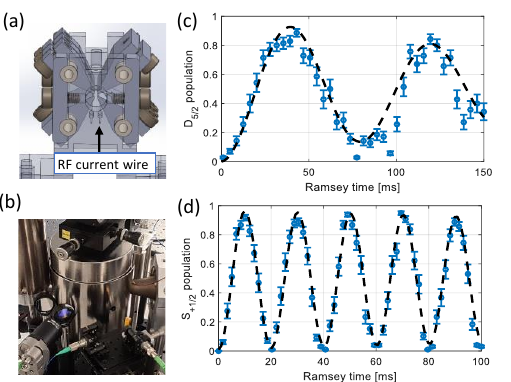}
    \caption{The ion trap and coherence times. (a-b) A compact vacuum chamber surrounded by a single-layer mu-metal shield accommodates a macroscopic ion trap. A current-carrying wire just a few mm from the ions allows the drive of magnetic dipole transitions, which are a central part of the DD scheme. (c) Ramsey spectroscopy measurement on the clock transition, showing laser-ion coherence time of around 150 ms (d) Ramsey spectroscopy on the two Zeeman states of the S-manifold indicates a Zeeman coherence time of many 100’s of ms owing to the efficient magnetic shielding}
    \label{fig:figure1}
\end{figure} 

Only recently have the first demonstrations of multi-ion clocks appeared, employing different approaches. One circumvents the problem by picking an atomic species and transitions with negligible sensitivity to the main broadening mechanism \cite{keller2019controlling,steinel2023evaluation}. however, this choice may come at the expense of other properties, such as challenging wavelength and the need for two-species operation. Another possibility is to work close to the QPS nulling angle \cite{tan2019suppressing}. Here, controlling the direction of the magnetic field to the required precision can be very challenging. Dynamical decoupling (DD) techniques \cite{aharon2019robust, Shaniv2019QuadrupoleDecoupling,kaewuam2020hyperfine,pelzer2024multi}, which average over multiple Zeeman components, can effectively mitigate both the QPS and the LZS and provide a viable avenue to enhance the accuracy of multi-ion optical clocks.

\begin{figure*}[t]
    \center
    \includegraphics[trim={{0cm 6.2cm 0cm 0cm}},clip,width=17cm]{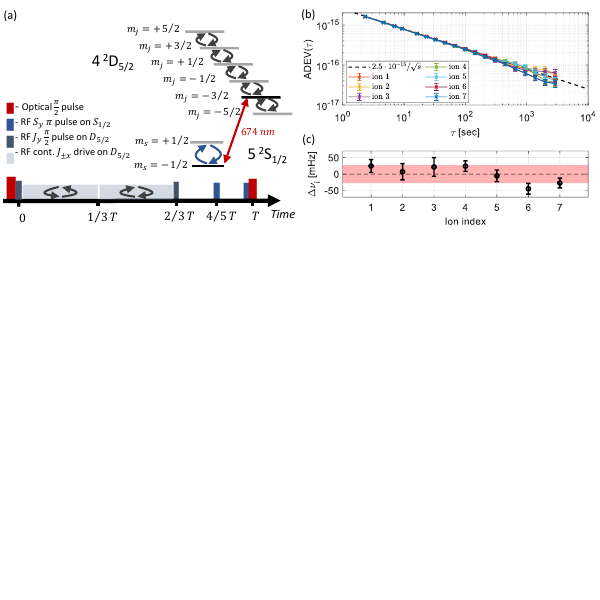}
    \caption{Suppressing QPS and ZS shift with QCDD Scheme. (a) A partial level-structure scheme of the $^{88}$Sr$^+$ ion that is relevant for the clock operation and schematics of the QCDD time sequence. (b) Overlapping Allan deviation analysis of a measurement comparing each of seven ions in a chain to their mean when applying the QCDD scheme. A servo was applied to keep the laser in the linear part of the Ramsey fringe by taking the average result of all ions. The plotted results are extracted by the residual population difference between the ions with Ramsey interrogation time of 90 ms and 60\% contrast. (c) The frequency shift of each ion relative to the mean when averaging over the entire measurement time.} 
    \label{figure2}
\end{figure*}	

Dynamic decoupling has been extensively studied and implemented across various quantum platforms, including nuclear magnetic resonance \cite{hahn1950spin,andrew1958nuclear,bennett1995heteronuclear}, neutral atoms \cite{dorscher2020dynamical, bishof2013optical}, solid-state spins \cite{golter2014protecting, pham2012enhanced}, and trapped ions \cite{yeh2023robust, kotler2011single}. Its applications range from enhanced spectroscopy\cite{stark2018clock} and sensing\cite{shaniv2017quantum} to quantum memory\cite{wang2021single,laucht2017dressed,zhang2014protected,biercuk2009optimized,timoney2011quantum,de2010universal} and robust quantum gates\cite{viola1998dynamical,morong2023engineering,xu2012coherence,barthel2023robust,fang2022crosstalk,pokharel2018demonstration}. The basic principle of dynamic decoupling relies on the non-commuting nature of different interactions. By applying a sequence of carefully timed and tailored control pulses, unwanted interactions between the quantum system and its environment can be averaged out, effectively isolating the system from external noise and errors. This method extends the coherence time of the quantum system, preserves quantum information for a longer duration, and improves performance.

Here, we experimentally investigate and characterize the performance of the QCDD scheme in a multi-ion clock setup. Following the proposal of R. Shaniv et al \cite{Shaniv2019QuadrupoleDecoupling}, we showcase the method's efficacy in suppressing the LZS and QPS on an optical transition in a crystal of up to seven $^{88}$Sr$^+$ ions. 

To thoroughly characterize the different systematic shifts associated with different clock interrogation methods, we employed a self-comparison approach. We evaluated the clock frequency by interlacing different interrogation methods throughout the integration time. This self-comparison enabled us to quantify systematic shifts arising from the interrogation schemes without needing another independent clock to compare with. Using self-comparison, we characterize the residual inhomogeneity between the ions. We study the systematic shift due to the RF drive used in our QCDD scheme. lastly, by self-comparison to single ion Ramsey interrogation, we show that the RF shift can be suppressed by interrogating two opposite transitions. These findings contribute valuable insights into the intricacies of multi-ion optical clocks, laying the groundwork for future advancements in precision measurement techniques and quantum technologies.

The QCDD technique encompasses a Ramsey-like spectroscopy scheme, composed of two $\pi/2$ optical pulses temporally separated before and after a radio-frequency (RF) sequence, which drives magnetic dipole transitions between the Zeeman states in a way that nullifies the QPS (concurrently mitigates other tensorial shifts, including the tensor ac-Stark shift) and LZS at each clock interrogation. A complete description of the analysis can be found in \cite{Shaniv2019QuadrupoleDecoupling}. Here, we only briefly state the underlying concept. For simplicity, we focus on the Hamiltonian of the six levels in D$_{5/2}$ clock excited manifold in the presence of near resonance RF drive. In the rotating wave approximation (RWA), The Hamiltonian takes the following form, 
\begin{equation}
\label{RWA}
H_{D}/\hbar=\Delta_BJ_z+Q_J(J^2-3J_z^2)+\Omega^D_{RF}(t)J_x. 
\end{equation}
Here, $\Delta_B$ is the detuning of the RF drive frequency from the Zeeman splitting due to a DC magnetic field. The second term accounts for the QPS with $Q_J$ containing the atomic level’s quadrupole moment, the gradient of the electric field, and geometric factors \cite{itano2000external}. Lastly, $\Omega^D_{RF}(t)$ is the time-dependent Rabi frequency that couples the various Zeeman levels through magnetic dipole interaction. When  $\Omega^D_{RF}(t)=0$ the time evolution operator is simply:
\begin{equation}
U_0(t)=e^{i[\Delta_BJ_z + Q_J(J^2-3J_z^2)]t}.
\label{free_time_evolution}
\end{equation}
However, if $\Omega_{RF}(t) >> Q_J,\Delta_B$, the time evolution operator can be written as,
\begin{equation}
U_{RF}(t)=e^{i[Q_J(J^2-\frac{3}{2}(J_z^2+J_y^2)) +\Omega_{RF}J_x]t}.
\end{equation}
To obtain the above expression from Eq.\ref{RWA}, we have used that $J_z^2=1/2(J_z^2+J_y^2)+1/2(J_z^2-J_y^2)$ where the second term, which does not commute with the $J_x$ drive and averages out, is omitted and so is the $\Delta_BJ_z$ term. 

The full sequence is depicted in the lower part of Fig.2a and consists of three parts. During the Ramsey time, a continuous drive is applied for $t=2/3\tau$, where the drive sign is flipped halfway. Applying this part between two $\pi/2$ $J_y$ rotation pulses results in a time evolution operator,
\begin{equation}
\Tilde{U}_{RF}(2/3\tau)=e^{i[Q_J(\frac{2}{3}J^2-(J_y^2+J_x^2))]\tau}.
\end{equation}
For the last $\tau/3$ part of the sequence, the system freely evolves according to Eq.\ref{free_time_evolution}. At the end of the sequence, the total phase contribution from the quadruple terms is summed to zero. We are only left with an LZS from the last part of the free evolution, which should be added to the LZS of the ground state that has been ignored so far. We can then cancel the DC (time-independent) component of the total LZS by applying a  $\pi$ pulse to flip the spin in the S$_{1/2}$ ground state at the appropriate time and flip it again just before the second optical $\pi/2 $ pulse that closes the Ramsey sequence (this is an asymmetric echo pulse where the exact timing depends on the g-factors ratio, see supplementary of ref\cite{Shaniv2019QuadrupoleDecoupling}). Thus, the QCCD scheme is free of the QPS and LZS.


Our multi-ion optical clock setup consists of a compact vacuum chamber that accommodates a room-temperature hand-assembled linear Paul trap (see Fig.1a). A single current-carrying wire positioned a few mm from the ions serves as a near-field antenna to resonantly drive Zeeman transitions in both the ground state S$_{1/2}$ and the excited D$_{5/2}$ manifolds. We can achieve Rabi frequencies of up to 100 kHz. However, the typical drive remains within a few kHz range to meet our experiments' specific requirements. 

The vacuum chamber is enclosed by a single layer of 1 mm thick mu-metal magnetic shielding. A constant magnetic field bias of $B\approx3 $ Gauss along the ions crystal is generated by permanent magnets inside the mu-metal shield complemented by small current coils that compensate for a small magnetic field gradient and stabilize long-term bias field drifts as measured on the ions. Magnetic shielding is a critical prerequisite as our transitions are first-order Zeeman-sensitive. Our typical Zeeman coherence, presented in Fig.1d, is of the order of a second, which is significantly longer than the optical clock coherence, showcasing the effectiveness of the magnetic shield (The coherence without the shield is limited to a few ms only).

Our clock laser at 674 nm has a typical coherence time of 150 ms, as measured by a Ramsey experiment shown in Fig.1c The laser system is based on diode lasers and includes an ECDL pre-stabilized to a ULE cavity ($\mathcal{F} \approx 100,000$) without temperature stabilization. This cavity also acts as a narrow optical filter, as only the transmitted light is utilized after it is amplified through injection locking to another bare diode laser. A second stabilization stage, which sets the laser's final performance, relies on locking to a frequency comb, which in turn is stabilized to a narrow Ti:Sapphire laser at 729 nm. The 729 nm laser is locked to another high-finesse ($\mathcal{F} \approx 300,000$) and thermally stabilized ULE cavity. 
All other lasers in the system are stabilized to a wavelength meter. 

Individual ion state detection within the chain is achieved by imaging the ions on an electron-multiplying charge-coupled device (EMCCD) camera with a 0.4 numerical aperture objective. We get high fidelity ($>99 \%$) state discrimination with less than one ms exposure time. Despite some overhead in the camera readout method, the clock interrogation time dominates our measurement duty cycle, which is around 70\%.
	
We begin by characterizing the performance of our QCDD scheme by comparing the frequencies of different ions in the crystal. To this end, we implement our scheme on a seven-ion crystal. We note here that the clock transition QPS varies by 10's of Hz between the ions (Our main clock beam is along the magnetic field and trap axial direction, which maximizes the QPS). Therefore, performing parallel narrow Rabi-type interrogation on all ions using a global beam is impossible. Moreover, a standard Ramsey scheme will result in an arbitrary fringe phase per ion, thus impeding their optimal usage. In contrast, our QCDD suppressed the above inhomogeneity, allowing us to exploit all the ions fully. To characterize the effectiveness of our QCDD scheme, we use individual ions measurements to estimate the ion-specific detuning for all ions in the crystal. The frequency shift per ion is extracted from the population imbalance of the two sides of the Ramsey fringe while accounting for the reduced fringe contrast of 60(10)\%, which was calibrated independently. The laser was kept on resonance by a feedback loop, with the error signal being the mean population of all the ions.

Figure 2.b presents an Allan deviation analysis of such measurement, comparing each of the seven ions to their mean. The Ramsey interrogation time in this measurement was 90 ms. Our analysis exhibits well-behaved shot-noise limited averaging. In addition, the horizontal red-shaded area indicates the standard deviation of the seven ions' detuning. This result shows that the QCDD scheme suppresses the QPS inhomogeneity to a level below 0.03 Hz or fractional frequency of $7\cdot10^{-17}$. Although these results are close to the level of the statistical error,
additional inhomogeneous frequency shifts may contribute to the measurement, for example, second-order Doppler and ac-Stark shifts due to inhomogeneous micromotion since, in this work, the trap was not operated at the magic RF frequency for $^{88}$Sr$^+$ \cite{dube2014high}. Thus, the variation should be considered as an upper bound.        

\begin{figure}[t]
    \includegraphics[trim={0cm 0cm 0cm 0cm},clip,width=9 cm]{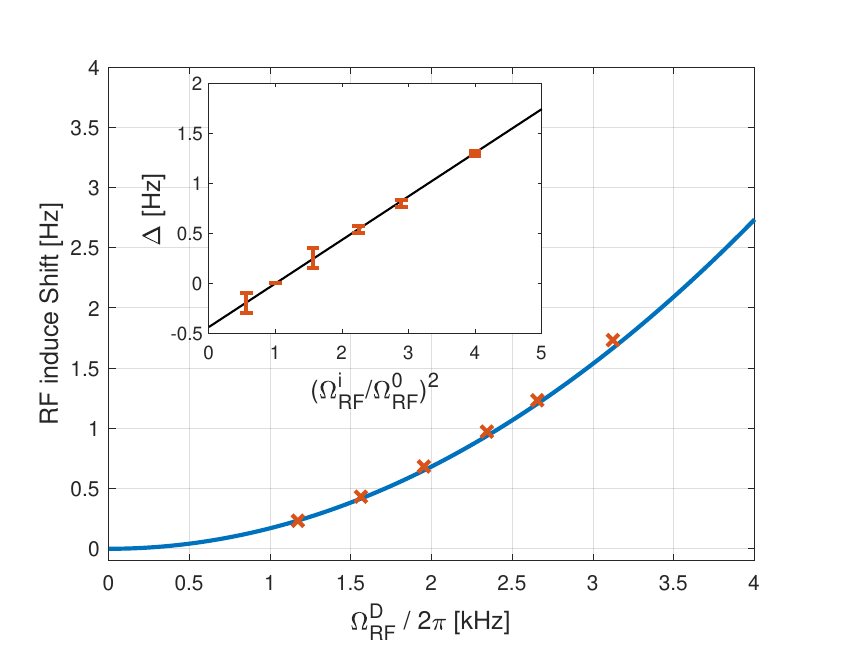}
    \caption{Systematic shift due to the RF drive during the QCDD scheme. Measurements of the RF-induced shift as a function of the drive strength in terms of its Rabi frequency. The inset presents the frequency shift $\Delta$ extracted from differential measurements with different Rabi frequencies as a function of the Rabi frequencies ratio squared. The black line in the inset is a linear fit of the measured data. In the main figure the red x's are the RF-induced shift, given the model $\delta_{RF}/2\pi=\alpha\cdot(\Omega^D_{RF}/2\pi)^2$ with  $\alpha=1.75(1)\cdot10^{-7}$ Hz$^{-1}$] obtain from of the fit in the inset. The blue solid line is theory without any fit parameter accounting for the ac-Stark shift.}
    \label{figure3}
\end{figure}

The measurement above is similar to the correlation spectroscopy used in \cite{Shaniv2019QuadrupoleDecoupling}. It is only sensitive to differential shifts between the ions and blind to any common frequency shifts, like those expected to arise from the RF pulses in the QCDD scheme. Thus, such a measurement alone does not prove the performance of the QCDD method in a realistic clock operation. A prominent shift is the cross-coupling to the S$_{1/2}$ manifold when driving on-resonance magnetic transition in the D$_{5/2}$. Because of the different g-factor in the D and S manifolds, driving on-resonance Zeeman transition in the D is accompanied by an off-resonance drive between the two spin states in the S, resulting in an ac-stark shift of the two ground states with respect to the D levels, hence shifting the clock transition.  

We measured the systematic shift due to the QCDD drive by varying the drive amplitude in an interlaced self-comparison measurement.  In this measurement, we ran a series of two interleaved QCDD interrogations. One was with a fixed drive strength of $\Omega^0_{RF}/2\pi=1.6$kHz used as a reference, while in the second, the drive amplitude $\Omega^i_{RF}$ was varied. Two independent servo loops were employed and averaged until the frequency difference between the two interleaved QCDD interrogations was evaluated with sufficient precision. The inset of Fig.\ref{figure3} shows the frequency shift as a function of the ratio of the drive Rabi frequencies squared $(\Omega^i_{RF}/\Omega^0_{RF})^2$. A straight line (black) is fitted to the measured data points with excellent agreement. We then use the fit parameters and a model of a pure quadratic shift in $\Omega^D_{RF}$ to re-plot (main figure) the RF-induced shift $\delta_{RF}$ directly in terms of drive Rabi frequency $\Omega^D_{RF}$, which gives $\delta_{RF}/2\pi=\alpha\cdot(\Omega^D/2\pi)^2$ with  $\alpha=1.75(1)\cdot10^{-7}$ Hz$^{-1}$. The blue solid line is a theory without any fit parameter accounting for the ac-stark shift of the $S_{1/2,+1/2}$ level due to the cross-coupling. The theory also includes the counter-rotating terms (correction to the RWA) due to the large detuning here. 

Even with a 2.5 kHz Rabi frequency used for DD, which results in about a 1 Hz shift and a conservative estimation of the stability of the drive amplitude to be at a level of 0.5 $\%$, we find that the contribution of the cross-coupling to the frequency uncertainty is below $10^{-17}$. In the following, we will show that this shift can also be mitigated.

 \begin{figure}[t!]
    \includegraphics[trim={0cm 0cm 0cm 0cm},clip,width=9 cm]{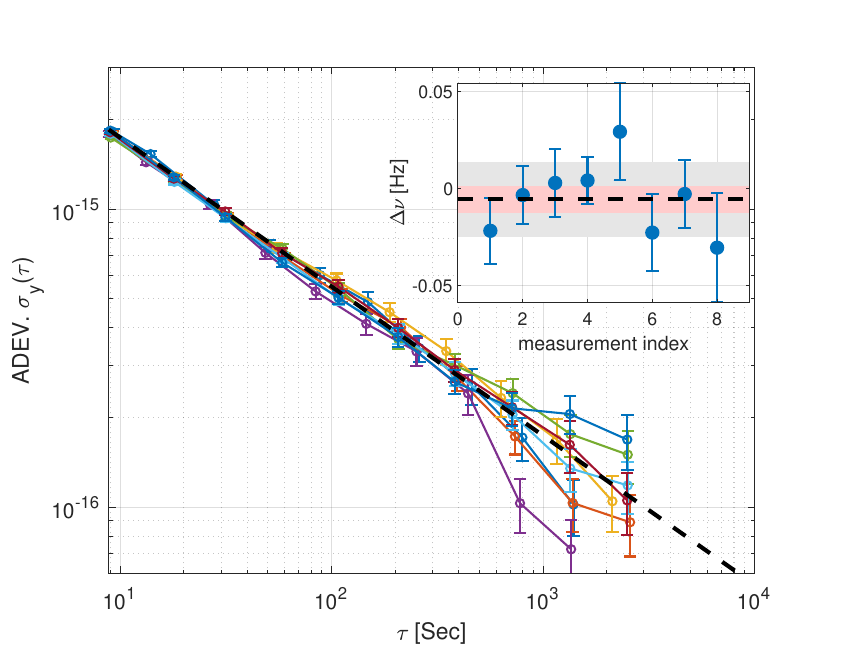}
    \caption{Self-comparison of five-ion crystal interrogated with QCDD to a single-ion Ramsey spectroscopy. The main figure presents an overlapping Allan deviation of the frequency difference between the two independent but interlaced measurements. The inset shows the absolute frequency difference $\Delta_T$ in each measurement. The gray shaded area indicates the standard deviation of the eight measurements.  Their average value (black dash line) and its uncertainty (red shaded area) obtained from the entire series of measurements is $-5(\pm 7)$ mHz.}
    \label{figure4}
\end{figure}	

While the RF-induced shift agrees well with theory and its magnitude is manageable in practice, it is always desired, if possible, to eliminate systematic shift by construction. Conveniently, this can be done with the RF induce shift by interrogating two opposite transitions with QCDD, for example, $S_{1/2,+1/2}\rightarrow D_{5/2,+3/2}$ and $S_{1/2,-1/2}\rightarrow D_{5/2,-3/2}$. Due to the symmetry of the shift, having a sign similar to the sign of the Zeeman states, it cancels out when taking the average of the transitions. Furthermore, evaluating the RF systematic shift in the scheme implemented here requires knowing the $g_S/g_D$ ratio with high precision. For $^{88}$Sr$^+$, this ratio has not been measured with high precision \cite{Barwood2012Characterization} and, as a result, limits the systematic shift uncertainty for a single transition interrogation. We note that a small modification to the DD scheme can also overcome this by adding two extra echo pulses that null the Zeeman shift in both manifolds independently. Nevertheless, the two-transition averaging has much more practical value and elegantly circumvents this difficulty.        

As a final experimental test, we investigate the absolute frequency difference between our two-transition QCDD scheme and a standard single-ion Ramsey interrogation scheme. This measurement uses a five-ion chain where a single measurement round consists of five different pairs of transitions/schemes interrogations. An interrogation pair comprises measuring each transition using $\delta=\pm 1/4\tau_m$ detunings, where $\tau_m$ is the interrogation time, in order not to be susceptible to variation in the optical $\pi/2$ pulses due to fluctuation in Rabi frequency. The QCDD part consists of measuring the two opposite transitions pairs $S_{1/2,+1/2}\rightarrow D_{5/2,+3/2}$ and $S_{1/2,-1/2}\rightarrow D_{5/2,-3/2}$. The results of these measurements are combined to form error signals for two servo loops, which, for convenience, are separated into the mean (for feedback on the laser frequency) and difference, which ideally should be constant. From the average of these two transitions, we can extract the laser detuning from the atomic transition $\delta_{QCDD}$, which is free of QPS and LZS:
\begin{equation}
\delta_{QCDD}= (f^+_{AOM} + f^-_{AOM})/2
\label{detuning_QCDD}
\end{equation}
Here,  $f^{\pm}_{AOM}$ is the frequency that is added to the laser using an AOM during the interrogation of the  $S_{1/2,\pm 1/2}\rightarrow D_{5/2,\pm 3/2}$ transitions. 

The above result is compared to the detuning $\delta_R$ obtained from the single ion Ramsey-like interrogations scheme, which consists of interrogating the three transitions: $S_{1/2,+1/2}\rightarrow D_{5/2,-1/2}$, $S_{1/2,+1/2}\rightarrow D_{5/2,+3/2}$ and $S_{1/2,+1/2}\rightarrow D_{5/2,+5/2}$. Averaging these three transitions nulls the QPS, where the LZS contribution is eliminated by adding two echo pulses (after $1/4\tau_m$ and $3/4\tau_m$) in both manifolds during the Ramsey time, which flips the Zeeman states. Each Ramsey transition is servo independently, where the error signal only accounts for the central ion in the crystal. $\delta_R$ is then obtained as follows,
\begin{equation}
\begin{split}
\delta_R= \frac{1}{3}\cdot( &f^{+5/2}_{AOM} +\frac{_1}{^2}f^S_{RF}-\frac{_5}{^2}f^D_{RF}\\
              + &f^{+3/2}_{AOM} +\frac{_1}{^2}f^S_{RF} -\frac{_3}{^2}f^D_{RF}\\ 
              + &f^{-1/2}_{AOM} +\frac{_1}{^2}f^S_{RF} +\frac{_1}{^2}f^D_{RF} ).
\end{split}
\label{detuning_ramsey}
\end{equation}
where  $f^{m_J}_{AOM}$ is the AOM frequency used to interrogate the transition $S_{1/2,+1/2}\rightarrow D_{5/2,m_J}$ and $f^{S(D)}_{RF}$ is the RF drive frequency used in the echo pulses applied to the S(D) manifold. The Ramsey servos were operated on top of the results obtained from the CQDD measurement. The reason is that the QCCD uses the entire five-ion chain and has lower projection noise. In this way, we benefit from the improved stability (reducing the probability of phase slip) without affecting the systematic shifts we are after. The choice of using echo pulses as an alternative to the more standard method of interrogating six transitions was due to technical limitations in our hardware.

%
The experiment results for the frequency difference between the two interrogation schemes  $\Delta=\delta_{QCDD}-\delta_R$ are presented in Fig.4. A series of eight measurements was taken over the course of a week and holds a total of 37 hours of acquisition time. The stability of the frequency difference $\Delta(\tau)$ is analyzed using the Allan deviation. The results exhibit shot noise-limited behavior throughout the averaging and down to $1\cdot 10^{-16}$(the maximal averaging time varies a little between experiments due to ion loss/laser unlocking, which results in aborting a measurement before it ends). The absolute frequency differences $\Delta_T$ of each measurement (average over the entire measurement time) are plotted in the inset (blue circle). The gray shaded area indicates the standard deviation of the eight measurements.  Their average value (black dash line) and its uncertainty (red shaded area) obtained from the entire series of measurements is $-5(\pm 7)$ mHz, and in terms of the fractional frequency uncertainty $-1(\pm 1) \cdot 10^{-17}$. 

In conclusion, we have studied and characterized a QCDD scheme to effectively suppress dominant frequency shifts in a multi-ion optical clock, specifically addressing the challenges of inhomogeneous frequency shifts due to QPS and LZS. We demonstrated more than three orders of magnitude suppression, resulting in inhomogeneity below $7\cdot10^{-17}$. By self-comparison to standard Ramsey interrogation and averaging two opposite transitions, we bound the systematic shift from the RF drive to be below  $2\cdot10^{-17}$. These results suggest that multi-ion clocks can achieve significant improvements in stability while maintaining high accuracy comparable to single-ion clocks.


We want to highlight the recent work by the PTB \cite{pelzer2024multi}, which also presents a multi-ion clock featuring continuous DD. Although both our approach and theirs are based on the same fundamental principle of using all Zeeman states to average out QPS and the LZS, there are some key differences in how they are implemented. Our method employs Ramsey spectroscopy with on-resonance RF QCDD, whereas the PTB approach uses off-resonance continuous mixing combined with Rabi spectroscopy. Additionally, our use of magnetic shielding significantly reduces the required bandwidth for DD and the resulting RF-induced shifts. This allows us to achieve and demonstrate high-accuracy performance (though only in self-comparison due to the absence of a reference).
      
This work was supported by the Israel Ministry of Science (IMOS grant 3-17376) and by the State of Lower Saxony Germany through joint research with the group of Piet O. Schmidt, with which we also had fruitful discussions.
\bibliography{references.bib}  
\end{document}